# First achievement of 100 million degree electron temperatures in a spherical torus: the EXL-50U

**Guang Yang,[1,2,3] Shaodong Song,[1,2,3] Songjian Li,[1,2,3] Xiaobing Xie,[1,2,3] Zhaokun Wang,[1,2,3] Xiang Gu,[1,2,3] Jia Li,[1,2,3] Xianli Huang,[1,2,3] Xinchen Jiang,[1,2,3,4] Yupeng Guan,[1,2,3] Huicong Ma,[1,2,3] Junquan Lin,[1,2,3,5] Dong Guo,[1,2,3,6] Jiaqi Dong,[7] Yunfeng Liang,[1,2,3] Yuejiang Shi,[1,2,3,*] and the EXL-50U Team**

[1]*Beijing ENN Fusion Energy Science and Technology Co., Ltd., Beijing 101111, China*
[2]*Beijing Key Laboratory of High Magnetic Field Spherical Torus Fusion Energy, Beijing 101111, China*
[3]*Hebei Key Laboratory of Compact Fusion, China*
[4]*School of Physics, East China University of Science and Technology, Shanghai 200237, China*
[5]*School of Resource Environment and Safety Engineering, University of South China, Hengyang 421001, China*
[6]*International Joint Research Laboratory of Magnetic Confinement Fusion and Plasma Physics, State Key Laboratory of Advanced Electromagnetic Engineering and Technology, School of Electrical and Electronic Engineering, Huazhong University of Science and Technology, Wuhan 430074, China*
[7]*Southwestern Institute of Physics, Chengdu 610041, China*
[*]*Correspondence: yjshi@ipp.ac.cn*

In future burning plasmas, alpha particles will primarily deposit their energy into electrons through collisional slowing-down, making electron temperature evolution and electron heat transport key factors in determining plasma performance.[1] Electron cyclotron resonance heating (ECRH) is expected to play an increasingly important role in future compact fusion devices,[2] which motivates the study of high-temperature plasmas sustained predominantly by electron heating. Spherical-torus (ST) experiments to date have predominantly relied on neutral beam injection, with most studies focusing on high-ion-temperature regimes.[3] Electron heat transport in high-electron-temperature, electron-heating-dominated ST plasmas therefore remains largely unexplored.

Here, we report an ST plasma in which Thomson scattering measures a central electron temperature exceeding 8.6 keV (100 million degrees kelvin) on the EXL-50U device.[4–7] The high-temperature phase is obtained at low density by sequential operation of two 50 GHz ECRH systems: ordinary-mode (O-mode) ECRH3 followed by extraordinary-mode (X-mode) ECRH4. In contrast to neutral-beam-heated high-ion-temperature ST scenarios, the discharge is sustained by ECRH only.

Magnetic equilibria are reconstructed with EFIT and the electron temperature and density profiles are taken from Thomson scattering.[8] Deposition of the X-mode ECRH4 wave is obtained from ray tracing on this equilibrium. In the absence of a measured effective charge, $Z_{eff}$ = 2.5 is assumed. This is a plausible value with intermittent diborane injection, but it remains a modeling uncertainty. The ion density then follows from quasi-neutrality. Nearby discharges (16873–16878) constrain the temperature ratio used in the modeling. On the same experimental day, X-ray crystal spectrometer[9] measurements give a core ion temperature clustered near 0.5 keV, which is adopted as the core boundary condition. The ion-temperature profile shape is assumed to follow the measured electron-temperature profile. At the present low collisionality, the inferred electron thermal diffusivity is insensitive to modest variations of this assumption. A steady-state electron heat balance is solved with the measured profiles and the ray-traced ECRH source. The effective electron thermal diffusivity is inferred from the resulting heat flux and compared with the neoclassical diffusivity from NCLASS.[10] Local linear gyrokinetic simulations with GENE[11] are performed at normalized radii ρ = 0.2 and ρ = 0.6.

Figure 1A shows the time history of discharge 18325 on EXL-50U (R = 0.6–0.8 m). In the high-temperature phase the discharge is sustained by 50 GHz electron cyclotron resonance heating at about 200 kW, with the magnetic axis held at R = 0.8 m so that Thomson scattering can access the core. The toroidal field at R = 0.6 m is 1.04 T and the plasma current is about 500 kA. Thomson scattering measures a central electron temperature of about 10 keV at 0.60 s, for which 8.6 keV is retained as a conservative lower bound because the error bars are large, at a core electron density of about $0.4 \times 10^{19}$ $m^{-3}$. No ion-temperature measurement is available in this high-temperature window. ECRH3 provides about 230 kW from the start of the discharge until 0.50 s and is then turned off. ECRH4 is applied at 0.50 s with about 200 kW and remains on through the high-temperature window. Hydrogen and diborane are injected alternately, with the puffing frequency minimized to keep a low density. From 0.42 to 0.60 s the core density remains near $0.4 \times 10^{19}$ $m^{-3}$. The core electron temperature remains only about 2 keV while ECRH3 is on, then rises promptly after the switch to ECRH4. The temperature peak lies near 0.60–0.65 s, and the high-temperature window lasts for roughly 150 ms. Because the injected power barely changes at the launcher switch, the temperature rise is attributed to a change in ECRH deposition rather than to an increase in heating power. The detailed mechanism responsible for the different deposition of the two launchers remains to be fully clarified. Whether the difference in launch polarization is the decisive cause of that deposition change remains open. The temperature collapses after about 0.65 s as the plasma current ramps down and the density rises.

At 0.60 s the electron temperature profile is strongly core-peaked (Figure 1B). The temperature falls to about 1 keV by $\rho \approx 0.5$, forming a steep gradient for $\rho \approx 0.2$–$0.6$. The density profile is much broader and only weakly peaked. The Thomson-scattering uncertainty in temperature is large for $\rho \lesssim 0.3$, as expected at this low density, and decreases toward the edge. A Maxwellian inversion is assumed. Possible non-thermal electrons under ECRH are not resolved by the present analysis and remain a caveat on the absolute central temperature. The contrast between a peaked temperature and a broad density indicates that the electron thermal energy is strongly localized near the axis.

Figure 1C shows that the core electron-to-ion temperature ratio decreases with core density and approaches unity at high density. At low density (below about $0.5 \times 10^{19}$ m$^{-3}$), the data concentrate in the high-ratio region, indicating strong electron–ion temperature decoupling. Discharge 18325 lies in this low-density branch. The red curve is a simple fit to the data.

Interpretive modeling of discharge 18325 at 0.60 s is summarized in Figure 1D. The driven current density and the electron heating power density are both localized within $\rho \lesssim 0.35$, with the main power peak near $\rho \approx 0.18$ (Figure 1D, panels (d1) and (d2)). This supports off-axis deposition of X-mode ECRH4. Over the entire minor radius the inferred diffusivity lies well above the neoclassical value, typically by two to four orders of magnitude, so the transport is anomalous (Figure 1D, panel (d3)). A local minimum of order 0.1 m$^2$ s$^{-1}$ occurs near $\rho \approx 0.05$. This minimum is still much larger than the neoclassical diffusivity, but it is low compared with the mid-radius and the edge. The electron heat flux rises from the axis, reaches a maximum of about 0.04 MW m$^{-2}$ near $\rho \approx 0.2$, and then decreases toward the edge as the flux-surface area increases (Figure 1D, panel (d4)). Off-axis heating together with this deep-core minimum is consistent with the strongly peaked temperature in Figure 1B.

Local linear GENE scans at two radii are shown in Figure 1E. At $\rho = 0.2$ the scan covers the ion-scale range $k_y\rho_s \lesssim 1$. The real frequency stays negative and decreases smoothly from about −0.45 to −1.05 $c_s/R_0$, with no branch crossing, so the mode propagates in the electron diamagnetic direction. The growth rate remains modest, about 0.13–0.17 $c_s/R_0$, with maxima near $k_y\rho_s \approx 0.4$ and 0.9. This ion-scale, electron-directed mode is consistent with a trapped-electron mode. At $\rho = 0.6$ the scan extends to $k_y\rho_s \approx 30$. The growth rate rises to a maximum of about 4 $c_s/R_0$ near $k_y\rho_s \approx 13$ and then falls, while the real frequency decreases almost linearly to about −33 $c_s/R_0$ and remains in the electron diamagnetic direction. The high-wavenumber peak together with a frequency proportional to wavenumber identifies an electron-temperature-gradient mode. The larger linear growth rate at $\rho = 0.6$ does not by itself imply a larger electron heat flux, because that mode has a much shorter wavelength than the ion-scale mode at $\rho = 0.2$. These linear growth rates are not a quantitative transport prediction.

EXL-50U has produced an ultra-hot, ECRH-dominated electron regime in a compact spherical torus. The central electron temperature exceeds 8.6 keV for about 150 ms at low density, with off-axis electron-cyclotron deposition, anomalous but comparatively low core electron heat transport, and linear microinstabilities in the electron diamagnetic direction: an ion-scale mode consistent with the trapped-electron mode at $\rho = 0.2$, and a strongly unstable electron-temperature-gradient mode at $\rho = 0.6$.

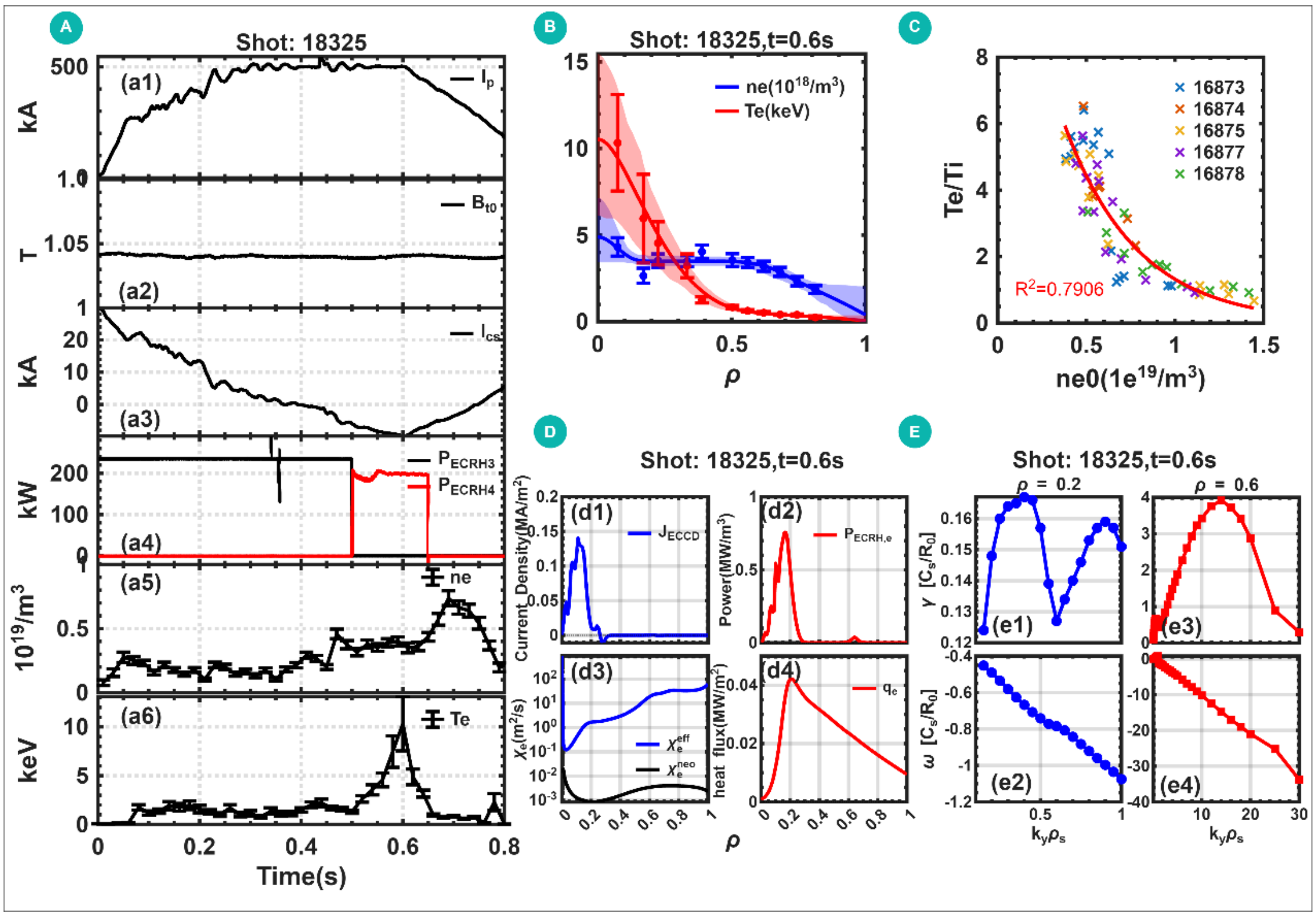

**Figure 1. High-electron-temperature discharge on the EXL-50U spherical torus.** (A) Time history of discharge 18325, panels (a1)–(a6): plasma current, toroidal field $B_{t0}$, central-solenoid current, injected power of O-mode ECRH3 and X-mode ECRH4, and core electron density and temperature from Thomson scattering. (B) Electron temperature (red) and density (blue) at 0.60 s. (C) Core electron-to-ion temperature ratio versus core electron density for discharges 16873–16878; at low density the points concentrate at a high ratio, and the red curve is a simple fit. (D) Interpretive modeling at 0.60 s under X-mode ECRH4: (d1) electron-cyclotron current-drive density, (d2) ECRH power-deposition density, (d3) effective electron thermal diffusivity compared with the neoclassical prediction, and (d4) electron heat flux. (E) Local linear GENE scan of $k_y\rho_s$ at 0.60 s, panels (e1)–(e4). Left, $\rho$ = 0.2; right, $\rho$ = 0.6. Upper panels are growth rates and lower panels are real frequencies, in units of $c_s/R_0$.

## FUNDING AND ACKNOWLEDGMENTS

This work was supported by ENN Science and Technology Development Co., Ltd. The authors thank the EXL-50U team for diagnostic and operational support. The funders had no role in study design, data collection and analysis, decision to publish, or preparation of the manuscript.

## AUTHOR CONTRIBUTIONS

All authors contributed to the manuscript and approved the final version.

## DECLARATION OF INTERESTS

The authors declare no conflicts of interest.